\documentclass[11pt]{article}

\usepackage{graphicx}

\title{Coherent and incoherent dynamic structure functions of the free
Fermi gas}

\author{F.Mazzanti$^1$, A.Polls$^1$ and J.Boronat$^2$  \\  \\
          $^1$ {\small Departament d'Estructura i Constituents
          de la Mat\`eria,} \\
          {\small Diagonal 645, Universitat de Barcelona,} \\
          {\small E-08028 Barcelona, Spain} \\
          $^2$ {\small Departament de F\'{\i}sica i Enginyeria Nuclear,} \\
          {\small Campus Nord B4--B5,
          Universitat Polit\`ecnica de Catalunya,} \\
          {\small E-08028 Barcelona, Spain} }

\begin{document}

\maketitle

\begin{abstract}
  A detailed calculation of the coherent and incoherent dynamic
structure functions of the free Fermi gas, starting from their
expressions in terms of the one- and semi--diagonal two--body density
matrices, is derived and discussed. Their behavior and evolution with the
momentum transfer is analyzed, and particular attention is devoted to
the contributions that both functions present at negative energies.
Finally, an analysis of the energy weighted sum rules satisfied by
both responses is also performed. Despite of the
simplicity of the model, some of the conclusions can be extended to
realistic systems.
\\  \\  \\
PACS: 05.30.Fk, 61.12.Bt
\\  \\  \\
KEYWORDS: Dynamic structure function, free Fermi gas
\end{abstract}

\maketitle

\bigskip

\bigskip

  Recent measurements of the dynamic structure function $S(q,\omega)$ in
pure liquid $^4$He and $^3$He, and $^3$He--$^4$He mixtures at high and
low momentum transfer \cite{sos1},\cite{fak1},\cite{gly1}
have revived the interest in the theoretical
analysis of $S(q,\omega)$ on both boson and fermion systems
\cite{gly1},\cite{maz1},\cite{gly2},\cite{mom1}
It is well
known that at high momentum transfer $q$ the scattering simplifies and
only the incoherent
part of the response survives. Therefore, splitting $S(q,\omega)$ in
coherent and incoherent terms may help to understand its nature. The
main purpose of this letter is to analyze both responses for the free
Fermi gas at any value of $q$. In this case they can be
calculated exactly because the Hamiltonian is
easily diagonalized. Despite the simplicity of the model, it can be
useful to understand certain
features of the response in real systems like $^3$He.

  The dynamic structure function of a quantum system of $N$
identical particles at zero temperature is defined as the probability of
coupling the ground state to all states compatible with a
momentum and energy transfer $(q,\omega)$ after a density fluctuation
$\rho_q=\sum_{j=1}^N e^{i{\bf q}\cdot{\bf r_j}}$ takes place
\begin{equation}
S(q,\omega) = \frac{1}{N} \sum_{\{n\}} \left| \langle n \mid \rho_q \mid 0
\rangle \right|^2 \delta \! \left( E_n - E_0 - \omega \right)
\label{aa-1}
\end{equation}

The Fourier transform of $S(q,\omega)$ generates the density--density
correlation factor $S(q,t)$ \cite{lov1}, which is the sum of the
incoherent and coherent density responses \cite{mom1}. These functions
are defined as
\begin{eqnarray}
S_{inc}(q,t) & = & \frac{1}{N} \sum_{j=1}^N
\langle e^{-i{\bf q}\cdot{\bf r_j}} e^{iHt} e^{i{\bf q}\cdot{\bf r_j}}
e^{-iHt} \rangle
\label{a-1a}  \\
S_{coh}(q,t) & = & \frac{1}{N} \sum_{i \neq j}^N
\langle e^{-i{\bf q}\cdot({\bf r_i}-{\bf r_j})}
e^{-i{\bf q}\cdot{\bf r_j}} e^{iHt} e^{i{\bf q}\cdot{\bf r_j}}
e^{-iHt} \rangle
\label{a-1b}
\end{eqnarray}
where $H$ and ${\bf r_j}$ are the Hamiltonian and the position
operators, respectively. Notice that with the definition given in 
(\ref{a-1b}) only those terms where particles $i$ and $j$ are strictly
different contribute to the coherent response, in contrast to other 
common definition where $S_{coh}(q,\omega)$ is taken as the whole 
density response and thus coincides with (\ref{aa-1}).

  In a realistic interacting system, where the
potential entering in $H$ depends on the position of the
particles, $H$ is non--diagonal neither in momentum nor in
configuration space, and the evaluation of the expectation values
in Eqs.~(\ref{aa-1}), (\ref{a-1a}) and (\ref{a-1b}) becomes
a formidable problem. Different approximations have been deviced in
the past, attempting to discern which are the leading contributions
in the limits of high \cite{ger1},\cite{sil1} and low
\cite{bor1} momentum
transfer. Even though a considerable amount of information concerning
the dynamics of quantum liquids and
nuclear systems have been gatered using those methods, precise
knowledge of the complete spectrum of the Hamiltonian is required if
one seeks to calculate all three responses exactly.

One of the few systems for which all eigenvalues and eigenvectors of the
Hamiltonian are known is the free Fermi gas, where $H$ is diagonal in
the momentum representation. In this case, the action of the momentum
translation operators $\exp(i{\bf q}\cdot{\bf r_j})$ appearing in
Eqs.~(\ref{a-1a}) and (\ref{a-1b}) is straightforward
\begin{eqnarray}
e^{-i{\bf q}\cdot{\bf r_j}} e^{iHt} e^{i{\bf q}\cdot{\bf r_j}} e^{-iHt}
& = & e^{-i{\bf q}\cdot{\bf r_j}} \exp \left[ it \sum_{k=1}^N
\frac{p_k^2}{2m} \right] e^{i{\bf q}\cdot{\bf r_j}} \exp \left[ -it
\sum_{l=1}^N \frac{p_l^2}{2m} \right] \nonumber \\
& = & \exp \left[ it
\left( \frac{({\bf p_j}+{\bf q})^2}{2m} - \frac{p_j^2}{2m} \right)
\right] \nonumber \\ [2mm]
& = & \exp \left[ it\frac{q^2}{2m} \right]
\exp \left[ it\frac{{\bf q}\cdot{\bf p_j}}{m} \right] \label{b-1}
\end{eqnarray}
and results in an exponential operator whose argument is
linear in ${\bf p_j}$, thus yielding the position translation operator
of particle $j$. After projecting on a complete basis of states in
configuration space, the incoherent and coherent responses become
functions of the one- and the semi--diagonal two--body density matrices
\begin{eqnarray}
S_{inc}(q,t) & = &
e^{i\omega_q t} \frac{1}{\rho} \rho_1(vt) \; , \label{c-1a} \\ [2mm]
S_{coh}(q,t) & = & e^{i\omega_q t} \frac{1}{\rho} \int d{\bf r}
\rho_2({\bf r}, 0; {\bf r}+{\bf v}t,0) e^{i{\bf q}\cdot{\bf r}} \; ,
\label{c-1b}
\end{eqnarray}
where $\rho$ is the particle number density, ${\bf v}={\bf q}/m$ is the
recoiling velocity of particle $j$ and $\omega_q = q^2/2m = mv^2/2$ its
associated kinetic energy. Comparing to the $1/q$ series of the response
derived by Gersch and coworkers in the early 70's, one readily notices
that $S_{inc}(q,t)$ and $S_{coh}(q,t)$ are described by the first terms
of the incoherent and coherent expansions, respectively. Higher order
terms in the series are related to integrals of the potential and vanish
when describing the response of a free system. As a consequence, the
incoherent response of the free Fermi gas is entirely given by the
Impulse Approximation (IA).

  Due to the close relation between the one--body density matrix
$\rho_1(x)$ and the momentum distribution $n(k)$, $S_{inc}(q,t)$ can
be Fourier transformed to yield the standard representation of the IA
\begin{eqnarray}
S_{inc}(q,\omega) & = & \frac{\nu}{(2\pi)^3\rho} \int d{\bf k} \, n(k) \,
\delta \! \left( \frac{({\bf k}+{\bf q})^2}{2m} - \frac{k^2}{2m}
- \omega \right)
\label{d-1a}
\\  [2mm]
& = & \frac{\nu m}{4\pi^2 \rho q} \int_{\mid Y \mid}^\infty
k \, n(k) \, dk ,
\label{d-1b}
\end{eqnarray}
$Y = m\omega/q - q/2$ being the West scaling variable
and $\nu$ the
degeneracy of each single--particle state of definite momentum. The
momentum distribution of the free Fermi gas is a Heaviside step
function $n(k) = \theta(k_F-k)$ that allows for the occupation of
states up to the Fermi surface only. Therefore, the
integral in Eq.~(\ref{d-1b}) can be performed and the result, expressed
in terms of a new set of dimensionless variables $\tilde{q}=q/k_F$
and $\tilde{\omega}=\omega/\epsilon_F$ with $\epsilon_F = k_F^2/2m$ the
Fermi energy, is given by
\begin{equation}
S_{inc}(q,\omega) \equiv
\frac{1}{\epsilon_F}
S_{inc}(\tilde{q}, \tilde{\omega}) = \frac{1}{\epsilon_F}
\frac{3}{8\tilde{q}} \left[ 1 -
\frac{1}{4} \left( \frac{\tilde{\omega}}{\tilde{q}} - \tilde{q}
\right)^2 \right] \theta \! \left[ 1 - \frac{1}{2} \left|
\frac{\tilde{\omega}}{\tilde{q}} - \tilde{q} \right| \right]
\label{e-1}
\end{equation}
which defines the dimensionless incoherent response
$S_{inc}(\tilde{q},\tilde{\omega})$.

  Proceeding in a similar way, the coherent response of the free Fermi
gas can be brought to a form that closely resembles the IA
\begin{equation}
S_{coh}(q,\omega) = \frac{\nu}{(2\pi)^3 \rho} \int d{\bf k} \,
n({\bf k}, -{\bf q}) \delta \! \left( \frac{({\bf k}+{\bf q})}{2m} -
\frac{k^2}{2m} - \omega \right)
\label{g-1}
\end{equation}
where $n({\bf k},{\bf q}))$ is the generalized momentum distribution
introduced by Ristig and Clark
\begin{eqnarray}
n({\bf k},{\bf q}) \!\!\! & = & \!\!\! \frac{1}{\nu N} \rho \int d{\bf
r_1} d{\bf r'_1} d{\bf r_2} \,
\rho_2({\bf r_1}, {\bf r_2}; {\bf r_{1'}}, {\bf r_2})
\, e^{i{\bf k}\cdot({\bf r_1}-{\bf r'_1})}
\, e^{-i{\bf q}\cdot({\bf r_1}-{\bf r_2})}
\label{f-1a}   \\  [2mm]
& = & \!\!\! \frac{1}{\nu} \int d{\bf r} d{\bf r'} \, \rho_2({\bf r}, 0;
{\bf r'})
\, e^{-i{\bf k}\cdot({\bf r}-{\bf r'})} \, e^{-i{\bf q}\cdot{\bf r}}
\; .
\label{f-1b}
\end{eqnarray}

  Both the semi--diagonal two--body density matrix and the generalized
momentum distribution of the non--interacting system are easily derived
\begin{eqnarray}
\rho_2({\bf r_1}, {\bf r_2}; {\bf r'_1}, {\bf r_2})
\!\!\!\! & = & \!\!\!\! \rho^2
\left[ \ell(k_F r_{11'}) - \frac{1}{\nu} \ell(k_F r_{12})
\ell(k_F r_{1'2}) \right]
\label{h-1a}   \\
n({\bf k},{\bf q}) = \theta(k_F \!\!\!\! & - & \!\!\!\! k) \left[
(2\pi)^3 \rho \delta \! ({\bf q}) - \theta \! \left( k_F -
\| {\bf k} - {\bf q} \| \right) \right]
\label{h-1b}
\end{eqnarray}
where $\ell(z)$ is the statistical correlation function,
characteristic of fermion systems
\begin{equation}
\ell(z) = \frac{3}{z^3} \, \left[ \sin z - z\cos z \right] \; .
\label{i-1}
\end{equation}
and therefore, inserting $n({\bf k}, {\bf q})$ (\ref{h-1b}) in
Eq.(\ref{g-1}), the coherent response may be written as
\begin{equation}
S_{coh}(q,\omega) = \frac{\nu}{(2\pi)^3\rho} \int d{\bf k} \,
\theta \! (k_F-k) \left[ (2\pi)^3\rho \, \delta \!({\bf q}) -
\theta \! (k_F - \| {\bf k}-{\bf q} \| ) \right] \,
\delta \! \left( \frac{({\bf k}+{\bf q})^2}{2m} -
\frac{k^2}{2m} - \omega \right)
\label{i-2}
\end{equation}
The $\delta({\bf q})$ term in the integral contributes only at
$q\!=\!0$, while the other carries all the information at
finite values of the momentum transfer and can be evaluated analytically
\begin{equation}
S_{coh}(q,\omega) \equiv
\frac{1}{\epsilon_F} S_{coh}(\tilde{q},\tilde{\omega}) = \left\{
  \begin{array}{ll}
     \left.
     \begin{array}{ll}
       -\frac{1}{\epsilon_F} \frac{3}{8\tilde{q}} \left[
       1 - \frac{1}{4} \left( \frac{\tilde{\omega}}{\tilde{q}}
       - \tilde{q} \right)^2 \right]  &  \mbox{if $\,\,\,0 \geq
       \tilde{\omega} \geq \tilde{q}^2-2\tilde{q}$}   \\
       - \frac{1}{\epsilon_F} \frac{3}{8\tilde{q}} \left[
       1 - \frac{1}{4} \left( \frac{\tilde{\omega}}{\tilde{q}}
       + \tilde{q} \right)^2 \right]  &  \mbox{if $\,\,\,2\tilde{q}
       - \tilde{q}^2 \geq \tilde{\omega} \geq 0$}
     \end{array}
     \right\}   &   \mbox{for $\,\,\,\tilde{q} \leq 2$}   \\
    \,\,\,0   &   \mbox{for $\,\,\,\tilde{q}>2$}
  \end{array}
\right.
\label{j-1}
\end{equation}
thus defining the dimensionless coherent response
$S_{coh}(\tilde{q},\tilde{\omega})$.

Several conclusions on the behavior of the coherent and incoherent
responses can be drawn from Equations (\ref{e-1}) and (\ref{j-1}), some
of them being also valid for realistic interacting systems.

The incoherent response $S_{inc}(\tilde{q},\tilde{\omega})$ of the free
Fermi gas is positive defined for all energies between
$\tilde{q}^2-2\tilde{q}$ and $\tilde{q}^2+2\tilde{q}$, and vanishes out
of this range. At fixed $\tilde{q}$, $S_{inc}(\tilde{q},\tilde{\omega})$
is a quadratic polynomial in $\tilde{\omega}$ with its maximum located
at $\tilde{\omega}=\tilde{q}^2$, thus being symmetric around this
point. Hence, both $S_{inc}(\tilde{q},\tilde{\omega})$ and its
derivatives are continuous.

At momentum transfer $\tilde{q}<2$, $S_{coh}(\tilde{q},\tilde{\omega})$
is non--zero and negative in the range $\tilde{\omega} \in
(\tilde{q}^2-2\tilde{q},2\tilde{q}-\tilde{q}^2)$. It is split in two
different parts, one defined at negative energies and the other at
positive ones. At fixed $\tilde{q}$, both functions are quadratic
polynomials in $\tilde{\omega}$ that differ only in the sign of the
linear coefficient. This peculiarity gives rise to a symmetric and
continuous $S_{coh}(\tilde{q}, \tilde{\omega})$ that, at
$\tilde{\omega}=0$, presents both a minimum and a discontinuity in the
first derivative. At $q$'s greater than twice the Fermi momentum, the
coherent response vanishes.

The total response $S(\tilde{q},\tilde{\omega})$, which is the sum of
$S_{inc}(\tilde{q},\tilde{\omega})$ (\ref{e-1}) and
$S_{coh}(\tilde{q},\tilde{\omega})$ (\ref{j-1}), is the well known
Lindhard function \cite{lin1},\cite{pin1} that becomes totally
incoherent at $\tilde{q}>2$. At
$\tilde{q}$'s smaller than 2, both the coherent and incoherent responses
contribute and, even though they are of opposite sign in the free Fermi
gas, the total response remains always positive. This is a general
property of the dynamic structure function of all systems, as is
apparent from Eq.~(\ref{aa-1}). For instance, if at a certain energy one
of the two responses is negative, its absolute value has to be smaller
than the value of the other one at that point, which should be positive,
as this is the only way to produce a total $S(q,\omega)$ that is either
positive or zero.

This feature is particularly apparent at negative $\omega$'s, as the
energy conserving delta appearing in the definition of $S(q,\omega)$
(Eq.~(\ref{aa-1})) forces the non--zero contributions to appear at
positive energies only. The separation of the response in its coherent
and incoherent terms breaks this constrain and both $S_{inc}(q,\omega)$
and $S_{coh}(q,\omega)$ can locate part of their strength at $\omega<0$.
However, as the total response is zero in this range,
$S_{coh}(q,\omega)$ has to be equal to $-S_{inc}(q,\omega)$ at
$\tilde{\omega}<0$. This is once again a requirement for the response
derived directly from its definition, and so does not depend on the kind
of system under study. It is easy to check from equations (\ref{e-1})
and (\ref{j-1}) that this holds for the free Fermi gas.

Figures 1,2 and 3 show the total response of the free Fermi gas and its
incoherent and coherent parts at $\tilde{q}=0.01$, $\tilde{q}=1$ and
$\tilde{q}=1.9$. The upper plots show the Lindhard function (solid
line), while $S_{inc}(\tilde{q},\tilde{\omega})$ and
$S_{coh}(\tilde{q},\tilde{\omega})$ are drawn below (solid
and dashed lines, respectively). There are three main regions at
$\tilde{q}\!\leq\!2$ where the Lindhard function changes its behavior
due to the different $\tilde{\omega}$ dependence of
$S_{inc}(\tilde{q},\tilde{\omega})$ and
$S_{coh}(\tilde{q},\tilde{\omega})$. At negative $\tilde{\omega}$'s
lying between $\tilde{q}^2-2\tilde{q}$ and 0,
$S_{inc}(\tilde{q},\tilde{\omega})=-S_{coh}(\tilde{q},\tilde{\omega})$
and the total response vanishes. At positive energies smaller than
$2\tilde{q}-\tilde{q}^2$, the addition of the two responses cancel the
terms quadratic in $\tilde{\omega}$ (see Equations~(\ref{e-1}) and
(\ref{j-1})) and the total $S(\tilde{q},\tilde{\omega})$ is linear.
Finally, at $\tilde{q}^2+2\tilde{q} \geq \tilde{\omega} \geq
2\tilde{q}-\tilde{q}^2$ the coherent response vanishes and the dynamic
structure function becomes entirely incoherent and quadratic in
$\tilde{\omega}$. Out of these limits all three functions are zero.

The different behavior of $S_{inc}(\tilde{q},\tilde{\omega})$ and
$S_{coh}(\tilde{q},\tilde{\omega})$ is also reflected in figures 1--3.
At very low values of the momentum transfer, both
$S_{inc}(\tilde{q},\tilde{\omega})$ and
$S_{coh}(\tilde{q},\tilde{\omega})$ are large, but strong cancelations
produce the strength of the total $S(\tilde{q},\tilde{\omega})$ to be
drastically reduced. Notice that in this limit the region
$\tilde{\omega} \! \in \! (2\tilde{q}-\tilde{q}^2,
\tilde{q}^2-2\tilde{q})$ where $S(\tilde{q},\tilde{\omega})$ has only
incoherent contributions covers a range $\Delta \tilde{\omega} =
2\tilde{q}^2 \ll 1$, which is much smaller than the range where the two
responses coexist. As a consequence, $S(\tilde{q},\tilde{\omega})$ looks
linear in $\tilde{\omega}$ (see figure 1). When $\tilde{q}$ increases, as
seen in figure 2, both $S_{inc}(\tilde{q},\tilde{\omega})$ and
$S_{coh}(\tilde{q},\tilde{\omega})$ are spread and quenched, but the
former still shows a maximum well centered at
$\tilde{\omega}=\tilde{q}^2$ and the latter a minimum at
$\tilde{\omega}=0$. The minimum of the coherent response coincides with
the point where the two functions defined in Eq.~(\ref{j-1}) join
($\tilde{\omega}=0$), and the different value between the left and right
derivatives at that point produces a peak that sharpens with increasing
$\tilde{q}$. At $\tilde{q}=1$ the range where the coherent response is
non--zero becomes maximal, going from $\tilde{\omega}=-1$ to
$\tilde{\omega}=1$, but the incoherent response has a wider range of
existence, and the total $S(\tilde{q},\tilde{\omega})$ presents two
well differentiated regions, one linear and another quadratic in
$\tilde{\omega}$. As $\tilde{q}$ rises above this value,
$S_{coh}(\tilde{q},\tilde{\omega})$ reduces both its magnitude and its
range of definition and, in particular, at $\tilde{q}=1.9$ becomes much
smaller than $S_{inc}(\tilde{q},\tilde{\omega})$ (see figure 3), thus
producing a total $S(\tilde{q},\tilde{\omega})$ that is mostly
incoherent.

  Another useful tool in the study of the dynamic structure function are
the energy weighted sum rules, which can be also derived for
$S_{inc}(\tilde{q},\tilde{\omega})$ and
$S_{coh}(\tilde{q},\tilde{\omega})$. They are defined to be the moments
of the different responses
\begin{eqnarray}
m_{inc,(coh)}^{(\alpha)}(\tilde{q}) & = & \int_{-\infty}^\infty
d\tilde{\omega} \, \tilde{\omega}^\alpha
S_{inc,(coh)}(\tilde{q},\tilde{\omega})
\label{k-1a}  \\
m^{(\alpha)}(\tilde{q}) \equiv m_{inc}^{(\alpha)}(\tilde{q}) & + &
m_{coh}^{(\alpha)}(\tilde{q}) = \int_0^\infty d\tilde{\omega} \,
\tilde{\omega}^\alpha S(\tilde{q},\tilde{\omega})
\label{k-1b}
\end{eqnarray}
and, even though $m_{inc}^{(\alpha)}(\tilde{q})$ and
$m_{coh}^{(\alpha)}(\tilde{q})$ integrate over all energies,
$m^{(\alpha)}(\tilde{q})$, built from their sum, is an equivalent
integral of the total $S(\tilde{q},\tilde{\omega})$ but obviously
restricted to positive energies.

  The first moments read
\begin{eqnarray}
m_{inc}^{(0)}(\tilde{q}) & = & 1
\label{l-1a}  \\
m_{coh}^{(0)}(\tilde{q}) & = & S(q) - 1
\label{l-1b}  \\
m_{inc}^{(1)}(\tilde{q} & = & \tilde{q}^2
\label{l-1c}  \\
m_{coh}^{(1)}(\tilde{q}) & = & 0
\label{l-1d}
\end{eqnarray}
and are the same in a correlated system replacing $S(q)$ by the
appropriate static structure factor. For the free Fermi gas, all odd
moments of $S_{coh}(\tilde{q},\tilde{\omega})$ cancel due to the
symmetry of the coherent response around $\tilde{\omega}=0$, while even
ones are related to derivatives of $\rho_2$. Furthermore, as
$S_{inc}(\tilde{q},\tilde{\omega})$ is entirely given by the impulse
approximation, all odd order incoherent moments centered at
$\tilde{\omega}=\tilde{q}^2$ vanish.
Particle correlations in realistic
systems may cause the static structure factor to be greater or smaller
than 1 as a function of $q$, as happens in liquid helium. Even though
this change in the sign of $m_{coh}^{(0)}(q)$ can be obtained from a
$S_{coh}(q,\omega)$ symmetric around $\omega=0$ that has no nodes, it
seems more reasonable to think about a coherent response that has at
least one point where the function changes sign, located in such a
way that both $m_{coh}^{(0)}(q)$ and $m_{coh}^{(1)}(q)$ are
fulfilled. Therefore, this difference in the behavior of the zero
moment points towards a change in the structure of the coherent
response entirely induced by particle correlations.

  The contributions of
$S_{inc}(\tilde{q},\tilde{\omega})$ and
$S_{coh}(\tilde{q},\tilde{\omega})$ at $\tilde{\omega}<0$, moments
(\ref{l-1a}) to (\ref{l-1d}) differ from those obtained integrating
over positive energies only. In particular for the Fermi sea, the first
moments of $S_{inc}(\tilde{q},\tilde{\omega})$ integrated at
$\tilde{\omega}>0$ are
\begin{eqnarray}
\int_0^\infty d\tilde{\omega} \, S_{inc}(\tilde{q},\tilde{\omega})
& = & 1 - \frac{3}{2} \int_{\tilde{q}/2}^\infty \tilde{k}
\left( \tilde{k} - \frac{\tilde{q}}{2} \right) \, n(\tilde{k}) \,
d\tilde{k}
\label{m-1a}  \\
\int_0^\infty d\tilde{\omega} \, \tilde{\omega}
S_{inc}(\tilde{q},\tilde{\omega}) & = & \tilde{q}^2 +
\frac{3\tilde{q}}{4} \int_{\tilde{q}/2}^\infty \tilde{k} \left(
\tilde{k} - \frac{\tilde{q}}{2} \right)^2 n(\tilde{k}) \, d\tilde{k}
\label{m-1b}
\end{eqnarray}
and coincide with the previous ones at $\tilde{q}>2$ where
$S(\tilde{q},\tilde{\omega}) = S_{inc}(\tilde{q},\tilde{\omega})$
The residual terms appearing in
the right hand side of equations (\ref{m-1a}) and (\ref{m-1b}) are the
manifestation of the negative energy contributions, and are the same
in realistic systems when the incoherent response is described
by the impulse approximation. Notice that in such
systems the momentum distribution $n(k)$ extends up to infinity, and
so those terms contribute no matter how large $\tilde{q}$ is.
Nevertheless, $n(k)$ decreases rapidly with $k$ and they become
vanishingly small as $\tilde{q}$ increases.

  In summary, a direct calculation of the coherent and incoherent
density responses of the free Fermi gas has revealed significant
differences between them and with respect to the total dynamic
structure function. While the latter is known to be non--zero and
positive at $\omega > 0$ only, none of those constrains apply to
the coherent and incoherent responses separately. In particular, and
for the Fermi gas, $S_{coh}(q,\omega)$ is negative and symmetric
around $\omega=0$ and both $S_{inc}(q,\omega)$ and $S_{coh}(q,\omega)$
present non--negligible contributions at $\omega < 0$ that cancel
exactly once added up. This cancellation also holds in any realistic
system. This behavior is also reflected in the energy
weighted sum rules.

\vfill\eject


\vfill\eject


\section*{Figure captions}

\begin{itemize}

\item[\bf Figure 1.] Comparison between the dimensionless Lindhard
Function (solid line above) and the dimensionless incoherent and
coherent dynamic structure functions (solid and dashed lines below,
respectively) at $\tilde{q}=0.01$.

\item[\bf Figure 2.] Same as in figure 1 at $\tilde{q}=1$.

\item[\bf Figure 3.] Same as in figure 1 at $\tilde{q}=1.9$.

\end{itemize}

\pagebreak

\begin{figure}[!t]
\begin{center}
\includegraphics*[height=18cm]{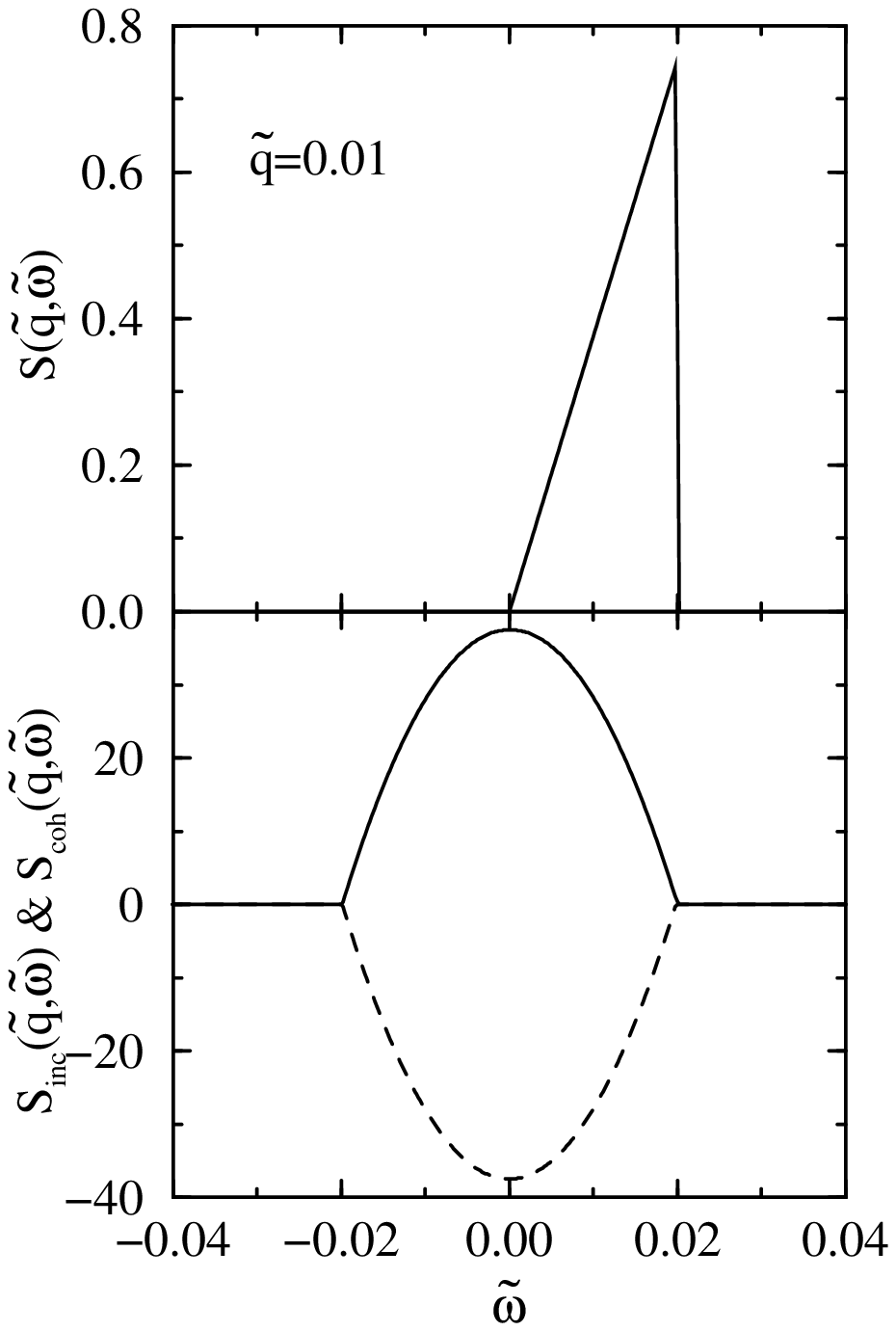}
\caption{}
\end{center}
\end{figure}

\pagebreak

\begin{figure}[!t]
\begin{center}
\includegraphics*[height=18cm]{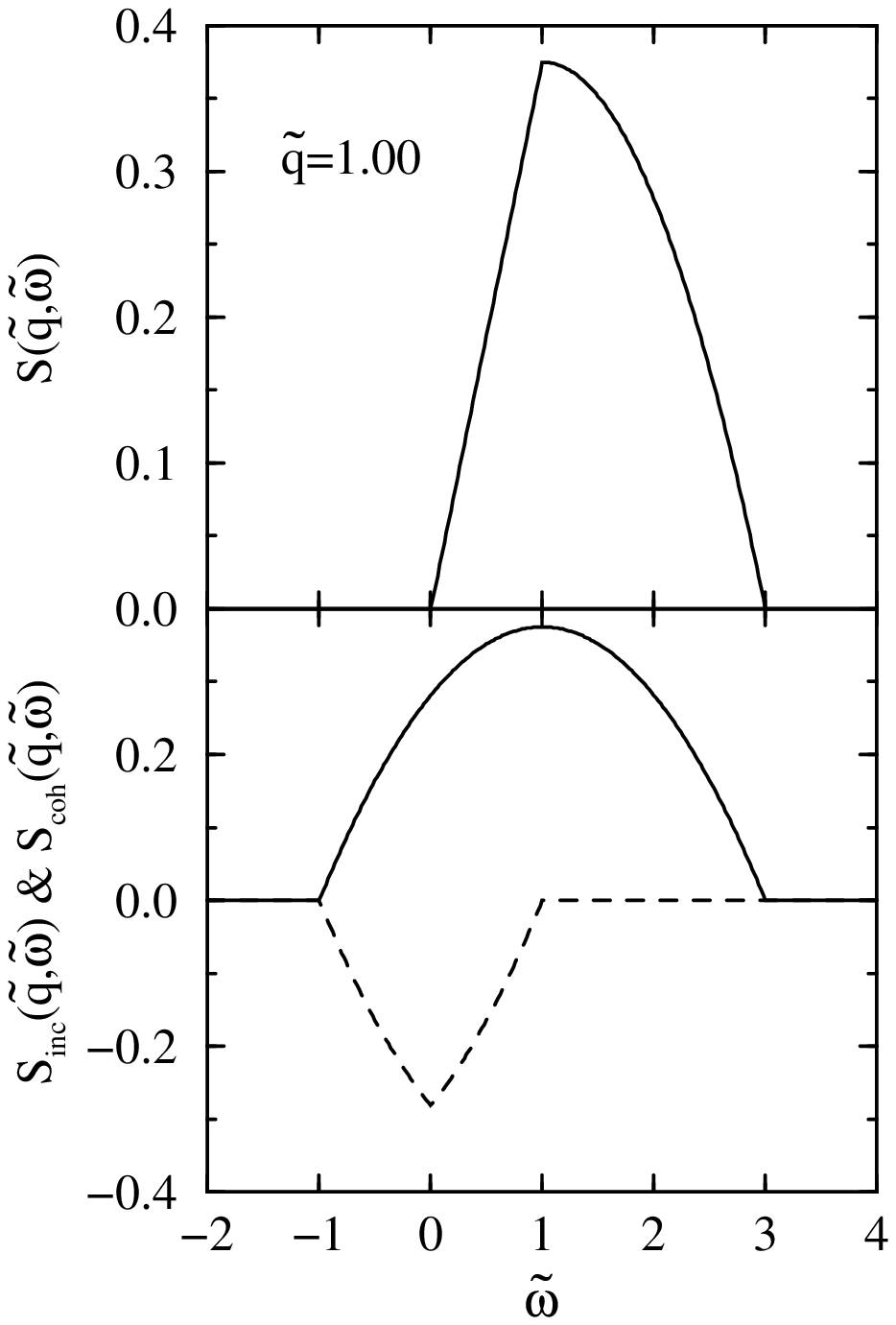}
\caption{}
\end{center}
\end{figure}

\pagebreak

\begin{figure}[!t]
\begin{center}
\includegraphics*[height=18cm]{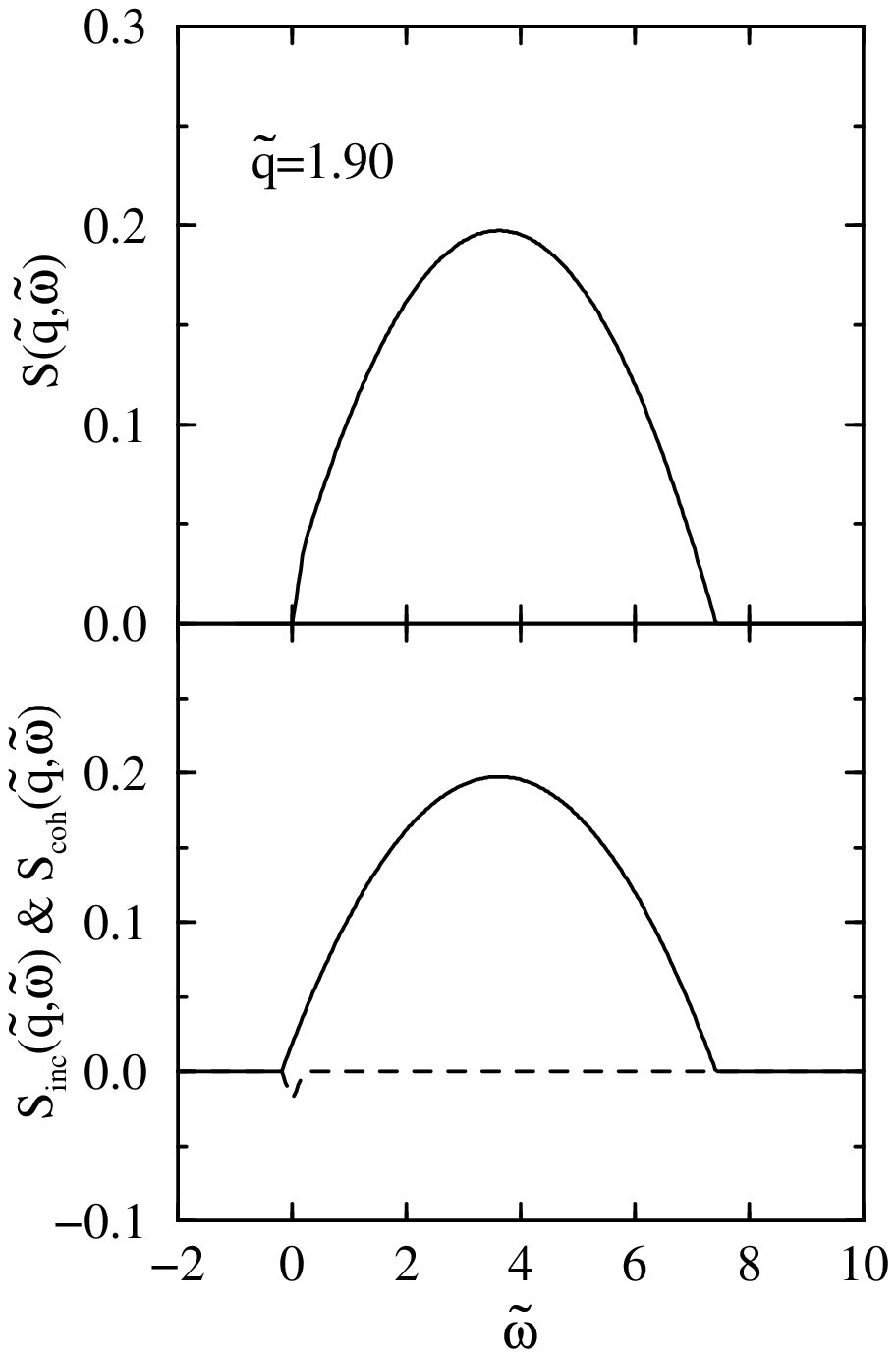}
\caption{}
\end{center}
\end{figure}

\end{document}